\documentclass[12pt,a4print]{article}
\usepackage{longtable}
\usepackage{layout}
\usepackage{amstex}

\begin{document}
\renewcommand{\theequation}{\thesection.\arabic{equation}}
\thispagestyle{empty}
\vspace*{-1.5cm}
\hfill {\small KL--TH 97/6} \\[8mm]

\message{reelletc.tex (Version 1.0): Befehle zur Darstellung |R  |N, Aufruf z.B. \string\bbbr}
%
%
\message{reelletc.tex (Version 1.0): Befehle zur Darstellung |R  |N, Aufruf z.B. \string\bbbr}
%
%
%
%
%
\font \smallescriptscriptfont = cmr5
\font \smallescriptfont       = cmr5 at 7pt
\font \smalletextfont         = cmr5 at 10pt
\font \tensans                = cmss10
\font \fivesans               = cmss10 at 5pt
\font \sixsans                = cmss10 at 6pt
\font \sevensans              = cmss10 at 7pt
\font \ninesans               = cmss10 at 9pt
\newfam\sansfam
\textfont\sansfam=\tensans\scriptfont\sansfam=\sevensans
\scriptscriptfont\sansfam=\fivesans
\def\sans{\fam\sansfam\tensans}
\def\bbbr{{\rm I\!R}} 
\def\bbbn{{\rm I\!N}} 
\def\bbbE{{\rm I\!E}} 
\def\bbbm{{\rm I\!M}}
\def\bbbh{{\rm I\!H}}
\def\bbbk{{\rm I\!K}}
\def\bbbd{{\rm I\!D}}
\def\bbbp{{\rm I\!P}}
\def\bbbone{{\mathchoice {\rm 1\mskip-4mu l} {\rm 1\mskip-4mu l}
{\rm 1\mskip-4.5mu l} {\rm 1\mskip-5mu l}}}
\def\bbbc{{\mathchoice {\setbox0=\hbox{$\displaystyle\rm C$}\hbox{\hbox
to0pt{\kern0.4\wd0\vrule height0.9\ht0\hss}\box0}}
{\setbox0=\hbox{$\textstyle\rm C$}\hbox{\hbox
to0pt{\kern0.4\wd0\vrule height0.9\ht0\hss}\box0}}
{\setbox0=\hbox{$\scriptstyle\rm C$}\hbox{\hbox
to0pt{\kern0.4\wd0\vrule height0.9\ht0\hss}\box0}}
{\setbox0=\hbox{$\scriptscriptstyle\rm C$}\hbox{\hbox
to0pt{\kern0.4\wd0\vrule height0.9\ht0\hss}\box0}}}}

\def\bbbe{{\mathchoice {\setbox0=\hbox{\smalletextfont e}\hbox{\raise
0.1\ht0\hbox to0pt{\kern0.4\wd0\vrule width0.3pt height0.7\ht0\hss}\box0}}
{\setbox0=\hbox{\smalletextfont e}\hbox{\raise
0.1\ht0\hbox to0pt{\kern0.4\wd0\vrule width0.3pt height0.7\ht0\hss}\box0}}
{\setbox0=\hbox{\smallescriptfont e}\hbox{\raise
0.1\ht0\hbox to0pt{\kern0.5\wd0\vrule width0.2pt height0.7\ht0\hss}\box0}}
{\setbox0=\hbox{\smallescriptscriptfont e}\hbox{\raise
0.1\ht0\hbox to0pt{\kern0.4\wd0\vrule width0.2pt height0.7\ht0\hss}\box0}}}}

\def\bbbq{{\mathchoice {\setbox0=\hbox{$\displaystyle\rm Q$}\hbox{\raise
0.15\ht0\hbox to0pt{\kern0.4\wd0\vrule height0.8\ht0\hss}\box0}}
{\setbox0=\hbox{$\textstyle\rm Q$}\hbox{\raise
0.15\ht0\hbox to0pt{\kern0.4\wd0\vrule height0.8\ht0\hss}\box0}}
{\setbox0=\hbox{$\scriptstyle\rm Q$}\hbox{\raise
0.15\ht0\hbox to0pt{\kern0.4\wd0\vrule height0.7\ht0\hss}\box0}}
{\setbox0=\hbox{$\scriptscriptstyle\rm Q$}\hbox{\raise
0.15\ht0\hbox to0pt{\kern0.4\wd0\vrule height0.7\ht0\hss}\box0}}}}

\def\bbbt{{\mathchoice {\setbox0=\hbox{$\displaystyle\rm
T$}\hbox{\hbox to0pt{\kern0.3\wd0\vrule height0.9\ht0\hss}\box0}}
{\setbox0=\hbox{$\textstyle\rm T$}\hbox{\hbox
to0pt{\kern0.3\wd0\vrule height0.9\ht0\hss}\box0}}
{\setbox0=\hbox{$\scriptstyle\rm T$}\hbox{\hbox
to0pt{\kern0.3\wd0\vrule height0.9\ht0\hss}\box0}}
{\setbox0=\hbox{$\scriptscriptstyle\rm T$}\hbox{\hbox
to0pt{\kern0.3\wd0\vrule height0.9\ht0\hss}\box0}}}}

\def\bbbs{{\mathchoice
{\setbox0=\hbox{$\displaystyle     \rm S$}\hbox{\raise0.5\ht0\hbox
to0pt{\kern0.35\wd0\vrule height0.45\ht0\hss}\hbox
to0pt{\kern0.55\wd0\vrule height0.5\ht0\hss}\box0}}
{\setbox0=\hbox{$\textstyle        \rm S$}\hbox{\raise0.5\ht0\hbox
to0pt{\kern0.35\wd0\vrule height0.45\ht0\hss}\hbox
to0pt{\kern0.55\wd0\vrule height0.5\ht0\hss}\box0}}
{\setbox0=\hbox{$\scriptstyle      \rm S$}\hbox{\raise0.5\ht0\hbox
to0pt{\kern0.35\wd0\vrule height0.45\ht0\hss}\raise0.05\ht0\hbox
to0pt{\kern0.5\wd0\vrule height0.45\ht0\hss}\box0}}
{\setbox0=\hbox{$\scriptscriptstyle\rm S$}\hbox{\raise0.5\ht0\hbox
to0pt{\kern0.4\wd0\vrule height0.45\ht0\hss}\raise0.05\ht0\hbox
to0pt{\kern0.55\wd0\vrule height0.45\ht0\hss}\box0}}}}

\def\bbbz{{\mathchoice {\hbox{$\sans\textstyle Z\kern-0.4em Z$}}
{\hbox{$\sans\textstyle Z\kern-0.4em Z$}}
{\hbox{$\sans\scriptstyle Z\kern-0.3em Z$}}
{\hbox{$\sans\scriptscriptstyle Z\kern-0.2em Z$}}}}
\setlength{\unitlength}{1cm}
\setlength{\topmargin}{-0.5cm}
\setlength{\textheight}{21cm}

\begin{center}
{\large\bf On the critical behaviour of hermitean $\pmb{f}$-matrix \\
models in the double scaling limit with $\pmb{f} \bf{ \ge 3}$}\\
\vspace{0.5cm}
{\large S. Balaska, J. Maeder and W. R\"uhl}\\
{\it Department of Physics, University of Kaiserslautern, P.O.Box 3049\\
67653 Kaiserslautern, Germany \\
E-mail: ruehl@@physik.uni-kl.de}\\
\vspace{5cm}
\begin{abstract}
An algorithm for the isolation of any singularity of $f$-matrix models in
the double scaling limit is presented. In particular it is proved by construction 
that only those universality classes exist that are known from $2$-matrix models.
\end{abstract}
\vspace{3cm}
{\it July 1997}
\end{center}
\newpage
\section{Introduction}
We investigate the critical behaviour of hermitean matrix models in the 
double scaling limit $N \to \infty$ and $g \to g_c$, where $N \times N$ is
the size of the matrices, $g$ is a coupling parameter, and a finite number 
$f$ of such random matrices is coupled to a chain. This scaling behaviour
describes two-dimensional quantum gravity coupled to the matter fields of
rational conformal field theories \cite{1}-\cite{3}. For $f = 1$ and $f = 2$ the
existing analysis of double scaling behaviour is complete \cite{4,5}. For
$f=3$ only a small number of examples are known \cite{6}, leading to just
three universality classes. We are going to present an algorithm which
allows the systematic construction of all double scaling limits for
$f \ge 3$. For an infinite subclass of $f=3$ models this algorithm employs
only linear algebra and all critical coefficients are rational. 
For all other cases of $f=3$ and in particular
for $f \ge 4$ we need to solve zeros of polynomials and the critical
coefficients turn into algebraic numbers. We emphasize that matrix models
are deeply related with Toda hierarchies \cite{7} but that this 
relationship has not been fruitful for the elucidation of the critical
behaviour.

In order to fix the notations we briefly describe hermitean matrix models.
The action is
\begin{eqnarray}
S(M^{(1)}, M^{(2)},...,M^{(f)}) = Tr \Big\{ \sum^f_{\alpha=1} V_{\alpha}
(M^{(\alpha)}) \nonumber \\
- \sum^{f-1}_{\alpha=1} c_{\alpha} M^{(\alpha)}M^{(\alpha+1)} \Big\}
\label{1.1}
\end{eqnarray}
where each potential $V_{\alpha}$ is a polynomial of degree $l_{\alpha}$
(only $l_{\alpha} \ge 3$ is of interest).
\begin{equation}
V_{\alpha}(t) = \sum^{l_{\alpha}}_{k=1} g^{(\alpha)}_k \frac{t^k}{k}
\label{1.2}
\end{equation}
Throughout the paper we assume that in relations between $l_1$ and $l_3$
w. l. o. g. $l_1>l_3$, if not stated otherwise.
Stability will be completely neglected in this work. The partition
function is
\begin{equation}
Z = \int \sum^f_{\alpha=1} dM^{(\alpha)} e^{-S}
\label{1.3}
\end{equation}
\begin{equation}
dM^{(\alpha)} = \prod_{\begin{array}{l} i \leq j \\ k < l \end{array}}
d(Re \, M^{(\alpha)}_{ij}) d(Im \, M^{(\alpha)}_{kl})
\label{1.4}
\end{equation}
The method of orthogonal polynomials \cite{8,9} makes use of
biorthonomal systems of polynomials
\begin{eqnarray}
\left\{ \Pi_m(\lambda), \; \tilde{\Pi}_m(\mu)\right\}^{\infty}_{m=0}
\nonumber \\
\lambda, \mu \in \bbbr
\label{1.5}
\end{eqnarray}
satisfying
\begin{eqnarray}
\int \prod^f_{\alpha=1} d\lambda^{(\alpha)} \Pi_m(\lambda^{(1)})
\tilde{\Pi}_n(\lambda^{(f)}) \nonumber \\
\times \exp \left\{ - \sum^f_{\alpha=1} V_{\alpha} (\lambda^{(\alpha)})
+ \sum^{f-1}_{\alpha=1} c_{\alpha} \lambda^{(\alpha)} \lambda^{(\alpha+1)} 
\right\} \nonumber \\
= \delta_{mn} \hspace{4cm}
\label{1.6}
\end{eqnarray}
Differentiation and multiplication matrices are introduced by
\begin{eqnarray}
\Pi^{\prime}_m &=& \sum _n(A_1)_{mn} \Pi_n \nonumber \\
\tilde{\Pi}^{\prime}_m &=& \sum_n (A_f)_{nm} \tilde{\Pi}_n
\label{1.7}
\end{eqnarray}
\begin{eqnarray}
\lambda \Pi_m (\lambda) &=& \sum _n(B_1)_{mn} \Pi_n (\lambda) \nonumber \\
\mu \tilde{\Pi}_m (\mu) &=& \sum_n (B_f)_{nm} \tilde{\Pi}_n (\mu)
\label{1.8}
\end{eqnarray}
With the help of auxiliary matrices
\[ B_2, B_3, ... B_{f-2} \]
we can derive Dyson-Schwinger equations
\begin{eqnarray}
A_1 + c_1 B_2 = V^{\prime}_1(B_1) \nonumber \\
c_{\alpha-1}B_{\alpha-1} + c_{\alpha} B_{\alpha+1} = V^{\prime}_{\alpha}
(B_{\alpha}) \nonumber \\
2 \leq \alpha \leq f-1 \nonumber \\
A_f + c_{f-1} B_{f-1} = V^{\prime}_f(B_f)
\label{1.9}
\end{eqnarray}
These matrices have support at
\begin{eqnarray}
(A_1)_{mn} = 0 & &\; \mbox{except for} \quad - \prod^f_{\alpha=1}(l_{\alpha}
-1) \leq n-m \leq -1 \nonumber \\
(A_f)_{mn} = 0 & &\; \mbox{except for} \quad 1 \leq n-m \leq  
\prod^f_{\alpha=1}(l_{\alpha} -1) 
\label{1.10}
\end{eqnarray}
and
\begin{equation}
(B_{\alpha})_{mn} = 0 \quad \mbox{except for} \quad - \prod_{\beta > \alpha}
(l_{\beta}-1) \leq n-m \leq \prod_{\beta < \alpha} (l_{\beta}-1)
\label{1.11}
\end{equation}
From (\ref{1.7}), (\ref{1.8}) follows
\begin{equation}
[B_1,A_1] = 1
\label{1.12}
\end{equation}
and the Dyson-Schwinger equations imply
\begin{eqnarray}
[B_1,A_1] = c_1[B_2,B_1] = c_2[B_3,B_2] = ... \nonumber \\
= c_{f-1}[B_f,B_{f-1}] = [A_f,B_f]
\label{1.13}
\end{eqnarray}
One can easily scale all the $\{c_\alpha\}$ to one in (\ref{1.9}), (\ref{1.13}).

For convenience we present a summary of our results at the end of this
work. However, we can anticipate that the universality classes found are
those and only those $[p,q]$ of the two-matrix model: $p$ and $q$ are either
coprime or coprime after a division by a factor $r$  different from
$p$ and $q$ \cite{5}. Then $p$ and $q$ are the orders of a pair of differential 
operators of a generalized Korteweg--de Vries hierarchy.

\setcounter{equation}{0}
\section{Solving the Dyson-Schwinger equations}
The solutions of the Dyson-Schwinger equations are obtained in two steps
perturbatively. One observes that any singularity obtained in the double scaling 
limit is already determined by the solution of the Dyson-Schwinger equations
at leading perturbative order. At this order we assume
\begin{equation}
(B_{\alpha})_{n,n+m} \begin{array}{l} \\ \longrightarrow \\ {\rm l.o.}
\end{array} \rho_m^{(\alpha)}
\label{2.1}
\end{equation}
so that for each $B_{\alpha}$ we obtain a generating function
\begin{equation}
r^{(\alpha)}(z) = \sum_m \rho_m^{(\alpha)} z^m
\label{2.2}
\end{equation}
with $z$ a complex variable (see (\ref{2.17})). The $\rho_m^{(\alpha)}$
are submitted to the constraints (\ref{1.11}) so that $r^{(\alpha)}(z)$ is
rational with poles only at $z=0$ and $z=\infty$. These generating
functions are inserted into the ``internal'' Schwinger-Dyson equations 
$(f \ge 3)$
\begin{equation}
B_{\alpha-1} + B_{\alpha+1} = V_{\alpha}^{\prime}(B_{\alpha}), \;
2 \leq \alpha \leq f-1
\label{2.3}
\end{equation}
Each $r^{(\alpha)}(z)$ is required to exhibit a zero of order $\lambda_
{\alpha}$ at $z=1$. The ansatz that leads to a set of zeros $\{\lambda_
{\alpha}\}$ is called a ``maximal critical point'' if it fixes all
unknowns in the Schwinger-Dyson equations at leading order (no parameters
are left over). We determine only maximal critical points. The $\{\rho^
{(\alpha)}_m\}$ and the critical coupling constants (\ref{1.2})
\begin{equation}
\{g^{(\alpha)}_k\}, \; 2 \leq \alpha \leq f-1
\label{2.4}
\end{equation}
can then be calculated.
In the subsequent sections 3 (for $f=3$) and 4 (for $f \ge 4$) this
program is performed. The $\{\rho^{(\alpha)}_m\}$ are either rational or
algebraic irrational. In the case $f=3$ there is a whole subclass of
solutions where this program can be executed analytically resulting in
only rational solutions.

In the second step we consider the ``external'' Schwinger-Dyson equations
\begin{eqnarray}
A_1 + B_2 &=& V^{\prime}_1(B_1) \nonumber \\
A_f + B_{f-1} &=& V^{\prime}_f(B_f)
\label{2.5}
\end{eqnarray}
They are (at this perturbative order) only used to fix the remaining critical
coupling constants
\begin{equation}
\{g^{(\alpha)}_k\}, \quad \alpha = 1 \quad {\rm or} \quad \alpha = f
\label{2.6}
\end{equation}
This is done as follows. Denote the restriction of $r^{(\alpha)}(z)$ to
its holomorphic part at zero (infinity) by
\[ r^{(\alpha)}(z)_{\ge}, \quad (r^{(\alpha)}(z)_{\leq}) \] 
respectively. Then using (\ref{1.10}) we can reformulate (\ref{2.5}) as
\begin{eqnarray}
r^{(2)}(z)_{\ge} &=& V^{\prime}_1(r^{(1)}(z))_{\ge} \nonumber \\
r^{(f-1)}(z)_{\leq} &=& V^{\prime}_f(r^{(f)}(z))_{\leq}
\label{2.7}
\end{eqnarray}

Now from (\ref{1.11})
\begin{eqnarray}
r^{(1)}(z) &=& \rho^{(1)}_1z + r^{(1)}(z)_{\leq} \nonumber \\
r^{(f)}(z) &=& \rho^{(f)}_{-1} z^{-1} + r^{(f)}(z)_{\ge} \\
& &(\rho^{(1)}_1 \not= 0, \; \rho^{(f)}_{-1} \not= 0). \nonumber
\label{2.8}
\end{eqnarray}
We see that we can introduce new variables
\begin{eqnarray}
\tilde{{r}}_1 &=& r^{(1)}(z) \nonumber \\
\tilde{{r}}_f^{-1} &=& r^{(f)}(z)
\label{2.9}
\end{eqnarray}
by holomorphic maps in a neighborhood of $z=\infty \; (z=0)$, respectively,
consequently
\begin{eqnarray}
r^{(1)}(z)^k_{\ge} &=& \tilde{r}^k_1 + O(\frac{1}{\tilde{r}_1}) \quad (\tilde{r}_1 \to \infty)
\nonumber \\
r^{(f)}(z)^k_{\leq} &=& \tilde{r}^{-k}_f + O(\tilde{r}_f) \quad (\tilde{r}_f \to 0)
\label{2.10}
\end{eqnarray}
and by a series expansion we get
\begin{eqnarray}
z^n &=& \sum^n_{k=-\infty} a^{(1)}_{nk} (\tilde{r}_1)^k, \quad n \ge 0 \nonumber \\
z^n &=& \sum^{+\infty}_{k=n} a^{(f)}_{nk} (\tilde{r}_f)^k, \quad n \leq 0
\label{2.11}
\end{eqnarray}
The $\{a^{(\alpha)}_{nk}\}$ are rational functions of the $\{\rho^{(\alpha)}_m\}$
and can be calculated recursively. Then the desired critical coupling constants 
result from 
\begin{eqnarray}
g^{(1)}_{k+1} &=& \sum^{l_1-1}_{m=0} \rho^{(2)}_m a^{(1)}_{mk} \nonumber \\
g^{(f)}_{k+1} &=& \sum^0_{m=-(l_f-1)} \rho^{(f-1)}_m a^{(f)}_{mk}
\label{2.12}
\end{eqnarray}

Having performed the leading order solution, the perturbative expansion
proceeds as in the case $f=2$. We keep the critical coupling constants fixed,
multiply the whole action with
\[ \frac Ng \]
and tune
\begin{equation}
N \to \infty, \quad g \to g_c
\label{2.13}
\end{equation}
as follows.

The matrix labels $n, m$ become continuous in this ``double scaling'' limit
\begin{equation}
\frac nN = \xi, \quad 0 \leq \xi \leq 1
\label{2.14}
\end{equation}
We replace the label $N$ by the ``string coupling constant'' $a$ :
\begin{equation}
\frac 1N = a^{2-\gamma}, \quad \gamma < 0
\label{2.15}
\end{equation}
so that $a \to 0$ for $N \to \infty$. Moreover
\begin{equation}
\xi = \frac{g_c}{g}(1- a^2x)
\label{2.16}
\end{equation}
The variable $z$ (\ref{2.2}) is dual in the Fourier series sense to the
discrete matrix label $m$ (see (\ref{2.1})). Now we set
\begin{equation}
z = e^{i\varphi}
\label{2.17}
\end{equation}
\begin{equation}
\varphi = a^{-\gamma}p
\label{2.18}
\end{equation}
so that due to (\ref{2.16}), (\ref{2.17})
\begin{equation}
p = i \frac{d}{dx}
\label{2.19}
\end{equation}
is the quantum mechanical momentum operator corresponding to $x$. The
perturbative expansion is in powers of
\[ a^{-\gamma} \]
and $\gamma$ is defined at the end from the Korteweg-de Vries hierarchy
operators by
\begin{equation}
\gamma = \frac{-2}{p+q-1}
\label{2.20}
\end{equation}
In this limit the matrices $\{B_{\alpha}\}^f_{\alpha=1}$ become
differential operators $\{R_{\alpha}\}^f_{\alpha=1}$
\begin{equation}
B_{\alpha} \to a^{-\gamma\lambda_{\alpha}}R_{\alpha} \quad \alpha \in \{1,
\ldots,f\} 
\label{2.21}
\end{equation}
where the order of $R_{\alpha}$ is $\lambda_{\alpha}$. While the $f$-matrix
model belongs to a class denoted
\[ (l_1,l_2, \ldots, l_f) \]
we use square brackets
\[ [\lambda_1, \lambda_2, \ldots, \lambda_f] \]
to denote the critical points.

\setcounter{equation}{0}
\section{The three-matrix model}
For the three-matrix model we proceed as follows. We set
\begin{eqnarray}
(B_1)_{n,n+k} &=& r_k^{(1)}(n) \nonumber \\
(B_2)_{n,n+k} &=& r_k^{(2)}(n) \nonumber \\
(B_3)_{n,n+k} &=& r_k^{(3)}(n) 
\label{3.1}
\end{eqnarray}
and make the ansatz
\begin{eqnarray}
r_k^{(1)}(n-\frac12) &=& \rho_k^{(1)} + a^{-2\gamma}u^{(1)}_k(x) \nonumber \\
r_k^{(2)}(n-\frac12) &=& \rho_k^{(2)} + a^{-2\gamma}u^{(2)}_k(x) \nonumber \\
r_k^{(3)}(n-\frac12) &=& \rho_k^{(3)} + a^{-2\gamma}u^{(3)}_k(x)
\label{3.2}
\end{eqnarray}
By inserting this ansatz in the Dyson-Schwinger equations we get a perturbed 
system in powers of $a^{-\gamma}$.

To $0$-th order one has equations between the $\rho^{(1)}_k, \rho^{(2)}_k,
\rho^{(3)}_k$ and the coupling constants $g^{(1)}_k, g^{(2)}_k, g^{(3)}_k$.
First we will concentrate on this system of equations.

In order to solve them we make a more precise ansatz for the   
$\rho^{(1)}_k, \rho^{(2)}_k$ and $\rho^{(3)}_k$. We introduce the generating
functions
\begin{eqnarray}
r^{(1)}(z) &=& \sum^1_{m=-(l_2-1)(l_3-1)} \rho^{(1)}_mz^m \nonumber \\
r^{(2)}(z) &=& \sum^{(l_1-1)}_{m=-(l_3-1)} \rho^{(2)}_mz^m \nonumber \\
r^{(3)}(z) &=& \sum^{(l_2-1)(l_1-1)}_{m=-1} \rho^{(3)}_mz^m
\label{3.3}
\end{eqnarray}
and assume they are of the form
\begin{eqnarray}
r^{(1)}(z) &=& \frac{(z-1)^{\lambda_1}P_{(l_2-1)(l_3-1)+1-\lambda_1}(z)}
{z^{(l_2-1)(l_3-1)}} \nonumber \\
r^{(2)}(z) &=& \frac{(z-1)^{\lambda_2}P_{(l_1+l_3-2)-\lambda_2}(z)}
{z^{(l_3-1)}} \nonumber \\ 
r^{(3)}(z) &=& \frac{(z-1)^{\lambda_3}P_{(l_1-1)(l_2-1)+1-\lambda_3}(z)}{z} 
\label{3.4}
\end{eqnarray}
where the $P_n$ are polynomials of degree $n$.

Inserting such an ansatz in the Dyson-Schwinger equations of order zero
one obtains the values for the critical coupling constants $g^{(1)}_k,
g^{(2)}_k, g^{(3)}_k$ and can fix the coefficients of the polynomials
$P_n$.

In the case of the three-matrix model, in contrast to the
two-matrix model (compare \cite{6}), more than one maximal critical
point can be found for each model. For example the (4,3,3) three-matrix 
model has three different
maximal critical points. They will be discussed in detail in Appendix B.

Among the critical points exist in general two different types which will
be called type I and type II. Critical points of type I are given if
$\lambda_1 = \lambda_3$. Then we have $R_1 = -R_3$ and the commutators
$[R_2,R_1]$ and $[R_3,R_2]$ are obviously identical and therefore
compatible with (\ref{1.13}). For the critical points of type II we have
$\lambda_1 \not= \lambda_3$. If $\lambda_1 < \lambda_3$ then to leading
order $R_1$ and $R_2$ are
 identical or $R_1$ is a power of $R_2$ up to a multiplicative constant and
their commutator is zero. But in higher orders the commutator of $R_1$ and
$R_2$ is the same as the one between $R_2$ and $R_3$ and again $[R_2,R_1]$
and $[R_3,R_2]$ are compatible. Of course, in this case the r\^ole of
$R_1$ and $R_3$ can be exchanged.

The appearance of the two types can be understood from examining the
``internal'' Dyson-Schwinger equation (\ref{2.3}). We may choose the ansatz
such that to leading order $B_1$ and $B_3$ compensate each other (type I) and
$V^{\prime}_2(B_2)$ only contributes in  higher orders. Or to leading
order $B_1$ and $V^{\prime}_2(B_2)$ compensate and $B_3$ only is important
for higher order contributions (type II).

In particular with the ansatz
\begin{equation} \label{3.5} 
r^{(2)}(z) = \frac{(z-1)^{(l_1+l_3-2)}}{z^{(l_3-1)}} 
\end{equation}
which yields maximal order for the differential operator $R_2$, and for which 
the Dyson-Schwinger equations can be solved by linear algebra only (see
Appendix A), we obtain 
several interesting cases. If $(l_1-1)$ and $(l_3-1)$ have no common
divisor then we found a critical point of type $[l_2+1, \, l_1+l_3-2, \,
l_2+1]$. In the other cases, in which $n(l_1-1) = m(l_3-1), \; n,m \in
\bbbn \backslash \{0\}$ holds, we got critical points of higher order
than $[l_2+1, \, l_1+l_3-2, \, l_2+1]$ except in the cases where $(l_1-1) =
m(l_3-1), m \in \bbbn \backslash \{0,1\}$ and $l_2 > m$. In these cases
the construction of the maximal critical point fails which is shown in 
detail in Appendix A.

In the following  table we have listed some three-matrix models and their 
maximal critical points. Rational solutions are marked with the 
abbreviation ``rat''. Moreover we give as last entries in this table the
universality class $[p,q]$ of these critical points which were defined
first for two-matrix models.

\vspace{1cm}
\begin{tabular}{cccc}
$(l_1,l_2,l_3)$ \quad & $[\lambda_1,\lambda_2,\lambda_3]$ & 
$[\lambda_1,\lambda_2,\lambda_3]$ &  $[p,q]$ \\ 
&first type & second type & \\ \hline
4,3,3 & 4,5,4 rat. & & 5,4 \\
& & 4,4,5 & 5,4 \\
& & 3,3,6 & 7,3 \\
4,4,3 & 5,5,5 rat. & & 6,5 \\
& 6,3,6 & & 7,3 \\
& & 4,4,6 & 6,4 \\
& & 3,3,7 & 7,3 \\
4,5,3 & 6,5,6 rat. & & 6,5 \\
& & 4,4,7 & 7,4 \\
& & 3,3,8 & 8,3 \\
4,6,3 & 7,5,7, rat. & & 7,5 \\
& & 4,4,8 & 9,4 \\
& & 3,3,9 & 10,3 \\
4,7,3 & 8,5,8 rat. & & 8,5 \\
& & 6,3,9 & 10,3 \\
& & 4,4,9 & 9,4 \\
& & 3,3,10 & 10,3
\end{tabular}

Table 1: Examples of singularities and their universality classes of
three-matrix models with $(l_1,l_2,l_3) = (4,n,3)$
  
\setcounter{equation}{0}
\section{Four-matrix models and beyond}
If the model involves four or more matrices we must always start from an ansatz with
free parameters such as (four matrices)
\begin{equation}
r^{(2)}(z) = \frac{(z-1)^{\lambda_2}}{z^{(l_3-1)(l_4-1)}} \sum^{m_2}_{n=0} 
a_m^{(2)} z^m
\label{4.1}
\end{equation}
\begin{equation}
m_2 = (l_3-1)(l_4-1) + (l_1-1)-\lambda_2
\label{4.2}
\end{equation}
The two ``internal'' Schwinger-Dyson equations
\begin{eqnarray}
B_1 + B_3 &=& V^{\prime}_2(B_2) \nonumber \\
B_2+B_4 &=& V^{\prime}_3(B_3)
\label{4.3}
\end{eqnarray}
admit two compensation mechanisms of type I or one of type I and the
other of type II as illustrated in the four graphs: 

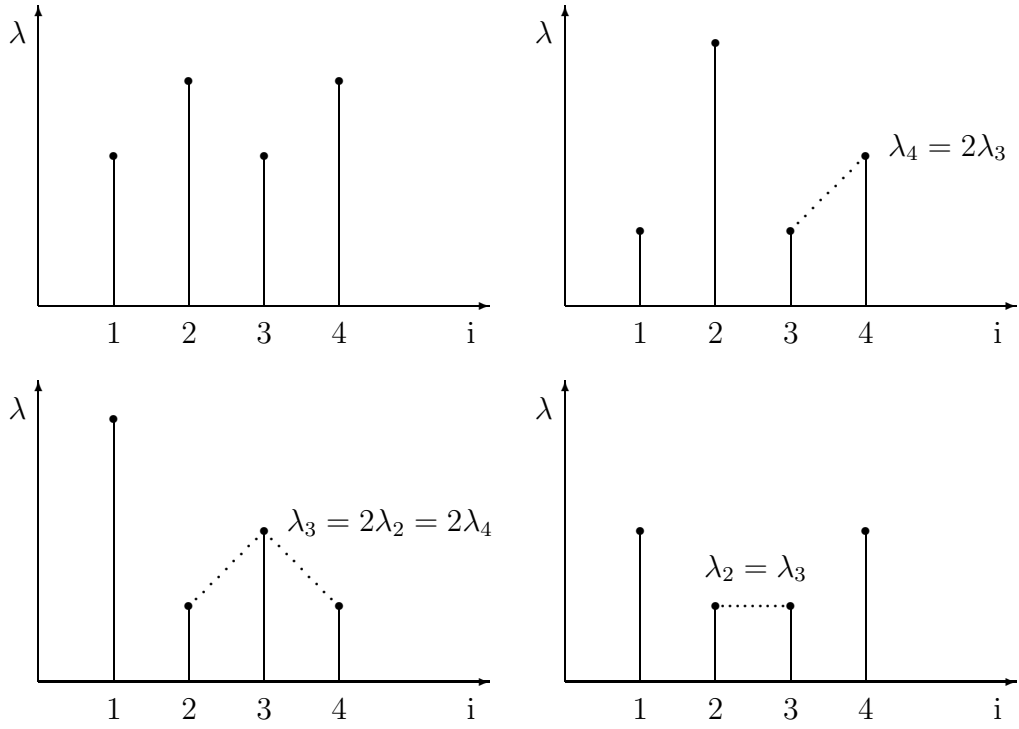
\begin{picture}(15,10)
\multiput(0.5,5)(7,0){2}{\vector(0,1){4}}
\multiput(0.5,5)(7,0){2}{\vector(1,0){6}}
\multiput(0.1,8.5)(7,0){2}{$\lambda$}
\multiput(6.2,4.5)(7,0){2}{i}
\multiput(1.4,4.5)(7,0){2}{1}
\multiput(2.4,4.5)(7,0){2}{2}
\multiput(3.4,4.5)(7,0){2}{3}
\multiput(4.4,4.5)(7,0){2}{4}
\put(1.5,5){\line(0,1){2}}
\put(1.5,7){\circle*{0.1}}
\put(2.5,5){\line(0,1){3}}
\put(2.5,8){\circle*{0.1}}
\put(3.5,5){\line(0,1){2}}
\put(3.5,7){\circle*{0.1}}
\put(4.5,5){\line(0,1){3}}
\put(4.5,8){\circle*{0.1}}
\put(8.5,5){\line(0,1){1}}
\put(8.5,6){\circle*{0.1}}
\put(9.5,5){\line(0,1){3.5}}
\put(9.5,8.5){\circle*{0.1}}
\put(10.5,5){\line(0,1){1}}
\put(10.5,6){\circle*{0.1}}
\put(11.5,5){\line(0,1){2}}
\put(11.5,7){\circle*{0.1}}
\multiput(10.5,6)(0.1,0.1){10}{\circle*{0.05}}
\put(11.8,7){$\lambda_4=2 \lambda_3$}
\multiput(0.5,0)(7,0){2}{\vector(0,1){4}}
\multiput(0.5,0)(7,0){2}{\vector(1,0){6}}
\multiput(0.1,3.5)(7,0){2}{$\lambda$}
\multiput(6.2,-0.5)(7,0){2}{i}
\multiput(1.4,-0.5)(7,0){2}{1}
\multiput(2.4,-0.5)(7,0){2}{2}
\multiput(3.4,-0.5)(7,0){2}{3}
\multiput(4.4,-0.5)(7,0){2}{4}
\put(1.5,0){\line(0,1){3.5}}
\put(1.5,3.5){\circle*{0.1}}
\put(2.5,0){\line(0,1){1}}
\put(2.5,1){\circle*{0.1}}
\put(3.5,0){\line(0,1){2}}
\put(3.5,2){\circle*{0.1}}
\put(4.5,0){\line(0,1){1}}
\put(4.5,1){\circle*{0.1}}
\multiput(2.5,1)(0.1,0.1){10}{\circle*{0.05}}
\multiput(3.5,2)(0.1,-0.1){10}{\circle*{0.05}}
\put(8.5,0){\line(0,1){2}}
\put(8.5,2){\circle*{0.1}}
\put(9.5,0){\line(0,1){1}}
\put(9.5,1){\circle*{0.1}}
\put(10.5,0){\line(0,1){1}}
\put(10.5,1){\circle*{0.1}}
\put(11.5,0){\line(0,1){2}}
\put(11.5,2){\circle*{0.1}}
\multiput(9.5,1)(0.1,0){10}{\circle*{0.05}}
\put(3.8,2){$\lambda_3=2 \lambda_2=2 \lambda_4$}
\put(9.35,1.4){$\lambda_2=\lambda_3$}
\end{picture}
\vspace{0.5cm}

Without doubt the universality classes are always $[p,q]$, this conclusion
is drawn solely from the possible compensation types. 

In the five-matrix-model
case the conclusion is the same, but the arguments are more general.
Namely, there are up to three types of compensations of type II possible 
(see the graph)

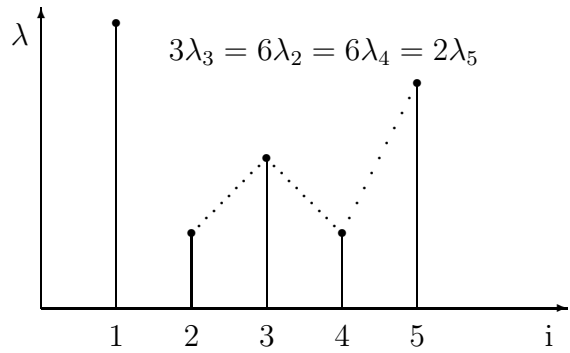
\begin{picture}(12,5)
\put(3.5,0){\vector(0,1){4}}
\put(3.5,0){\vector(1,0){7}}
\put(3.1,3.5){$\lambda$}
\put(10.2,-0.5){i}
\put(4.4,-0.5){1}
\put(5.4,-0.5){2}
\put(6.4,-0.5){3}
\put(7.4,-0.5){4}
\put(8.4,-0.5){5}
\put(4.5,0){\line(0,1){3.8}}
\put(4.5,3.8){\circle*{0.1}}
\put(5.5,0){\line(0,1){1}}
\put(5.5,1){\circle*{0.1}}
\put(6.5,0){\line(0,1){2}}
\put(6.5,2){\circle*{0.1}}
\put(7.5,0){\line(0,1){1}}
\put(7.5,1){\circle*{0.1}}
\put(8.5,0){\line(0,1){3}}
\put(8.5,3){\circle*{0.1}}
\multiput(5.5,1)(0.1,0.1){10}{\circle*{0.05}}
\multiput(6.5,2)(0.1,-0.1){10}{\circle*{0.05}}
\multiput(7.5,1)(0.1,0.2){10}{\circle*{0.05}}
\put(5.2,3.3){$3 \lambda_3=6 \lambda_2=6 \lambda_4=2 \lambda_5$}
\end{picture}
\vspace{1cm}

\noindent Of course, for a compensation one needs simply 
$\lambda_i = n \lambda_{i\pm1}, \; n \in \bbbn$, the explicit cases 
given here (as $\lambda_4 =2 \lambda_{3}$) are just examples.

One can imagine that what was shown with these examples holds true for 
all $f$-matrix models with arbitrary
$f$. Thus the universality classes are always of the two-matrix model type.
The actual calculation proceeds by solving the first Schwinger-Dyson
equation (\ref{4.3}) with the ansatz (\ref{4.1}) and the method of the
three-matrix model. Then $\{ \rho^{(1)}_m, \rho^{(3)}_m\}$ and $\{g^{(2)}_k
\}$ result as polynomials in $\{a^{(2)}_m \}$. Next we have to adjust the
 $\{a^{(2)}_m \}$ and the  $\{g^{(2)}_m \}$ so that
\begin{enumerate}
\item \begin{eqnarray}
r^{(4)}(z) &=& V^{\prime}_3(r^{(3)}(z)) - r^{(2)}(z) \nonumber \\
\rho^{(4)}_m &=& 0 \quad {\rm for} -(l_3-1)(l_4-1) \leq m \leq -2
\label{4.6}
\end{eqnarray}
\item $r^{(4)}(z)$ has a maximal zero of degree $\lambda_4$ at $z=1$.
\end{enumerate}
For simple four-matrix models this program can even be performed by hand,
e.g. for $(l_1, l_2, l_3, l_4) = (3,3,3,3)$.

In this case we find a 
\begin{equation}
[\lambda_1, \lambda_2, \lambda_3, \lambda_4] = [5,4,5,4]
\label{4.7}
\end{equation}
with
\begin{equation}
\mu_2 = 2, \quad g^{(2)}_3 = 1 \quad \mbox{(by normalization)}
\label{4.8}
\end{equation}
\begin{eqnarray}
\mu_3 = 1, \quad g^{(3)}_2 &=& \frac{3}{4\eta^3} (17 \eta^2 - 
24 \eta - 5) \nonumber \\
g^{(3)}_3 &=& \frac{1}{8\eta^2} (-\eta^2-6\eta + 15)
\label{4.9}
\end{eqnarray}
where $\eta$ is any of the three real solutions of
\begin{equation}
\eta^3 + 7\eta^2 - 15\eta - 5 = 0
\label{4.10}
\end{equation}
Examples how the different compensations, that we presented in this section 
actually occur, can be seen in appendix B.

\section{Summary}

An algorithm has been described by which all critical points of $f$-matrix
models with polynomial potentials in the double scaling limit $N \rightarrow
\infty$ can be derived. These critical points form universality classes $[p,q]$
where $p$ and $q$ are the orders of differential operators of a generalized
KdV hierarchy and satisfy only
\[
p \geq q, \quad p/q \notin \bbbn.
\]
We have given a new constructive argument to move this.

The algorithm treats the Dyson-Schwinger equations, derived from the orthogonal
polynomial approach, perturbatively by expansion in power series of a rational
power of $\frac{1}{N}$. For $f \geq 3$ these Dyson-Schwinger equations split
into $(f-2)$ ``internal'' and (two) ``external'' equations. A leading order
analysis of the internal equations fixes the critical point. The
external equations determine only the two external critical potentials. The
fact that only two differential operators result from $f$ matrices
$\{ B_i \}_{i=1}^{f}$ in the double scaling limit is the result of $f-2$
compensations. We distinguish between two different compensation mechanisms
(type I and type II) that act on any triplet $B_i, B_{i+1}, B_{i+2}$.
The differential operators of order $p$ and $q$ arise after the compensations
and at higer perturbative order. Thus we must be able to push the perturbative
order to any desired value. This fact was responsible already for the 
derivation of all critical points of the two-matrix models in \cite{6}
(including some not seen before) and is even more decisive for the analysis
of multi-matrix models. As usual the perturbative order of the commutator
\[
\, [B_1, B_2]
\]
is $p+q-1$ and gives the string susceptibility exponent $\gamma$ by
\[
\gamma = \frac{-2}{p+q-1}.
\]

In the case of the three-matrix models more detailed results have been 
obtained. We determined the maximal (i.e. parameter free) critical points and
found classes of ``rational'' and ``algebraic'' critical points, referring to
the values of their critical coupling constants. The rational class derived
from (\ref{3.5}) is absent however, if
\[
\frac{l_1-1}{l_3-1} \in \bbbn \quad \mbox{ or } \quad 
\frac{l_3-1}{l_1-1} \in \bbbn
\]
where $l_1, l_3$ denote the degree of the external critical potentials.

\newpage

\setcounter{equation}{0}
\begin{appendix}
\section{A linear algorithm}
We make the ansatz
\begin{equation}
r^{(2)}(z) = \frac{(z-1)^{l_1+l_3-2}}{z^{l_3-1}}
\label{A.1}
\end{equation}
\begin{equation}
\rho^{(2)}_m = (-1)^{l_1-m+1} {l_1+l_3-2 \choose m+l_3-1}
\label{A.2}
\end{equation}
We have the ``internal'' Schwinger-Dyson equation
\begin{equation}
r^{(1)}(z) + r^{(3)}(z) = V_2^{\prime}(r^{(2)}(z))
\label{A.3}
\end{equation}
Then (\ref{A.3}) amounts to
\begin{equation}
\sum^{l_2-1}_{k=1} g^{(2)}_k A_{m,k-1} = \left\{
\begin{array}{l}
\rho^{(2)}_m, \; -(l_2-1)(l_3-1) \leq m \leq -2 \\
\rho^{(1)}_m + \rho^{(3)}_m, \; - 1 \leq m \leq +1 \\
\rho^{(3)}_m, \; 2 \leq m \leq (l_1-1)(l_2-1)
\end{array} \right.
\label{A.4}
\end{equation}
with
\begin{equation}
A_{mk} = (-1)^{k(l_1-1)-m} {k(l_1+l_3-2) \choose k(l_3-1)+m}
\label{A.5}
\end{equation}
In order to produce zeros of order $\lambda_1, \lambda_3$ at $z=1$, the coefficients 
$\rho^{(1)}_m, \rho^{(3)}_m$ must fulfil
\begin{eqnarray}
\sum^{-(l_2-1)(l_3-1)}_{m=1} {m \choose n} \rho^{(1)}_m = 0, \quad
0 \leq n \leq \lambda_1-1 \nonumber \\
\sum^{(l_1-1)(l_2-1)}_{m=-1} {m \choose n} \rho^{(3)}_m = 0, \quad
0 \leq n \leq \lambda_3-1
\label{A.6}
\end{eqnarray}
where we want to maximize $\lambda_1, \lambda_3$.

With the shorthands
\begin{equation}
R^{(1,3)}_n = \sum^{+1}_{m=-1} {m \choose n} \rho^{(1,3)}_m
\label{A.7}
\end{equation}
the $m$-summations of (\ref{A.2}) can be done analytically to yield
\begin{eqnarray}
\sum^{l_2-1}_{k=0} g^{(2)}_{k+1} \Phi_{nk} = R^{(3)}_n \nonumber \\
\sum^{l_2-1}_{k=0} g^{(2)}_{k+1} \Psi_{nk} = R^{(1)}_n
\label{A.8}
\end{eqnarray}
with
\begin{eqnarray}
\Phi_{n,k} &=& (-1)^{K_1} {K_1+K_3 \choose K_3} \delta_{n0} \nonumber \\
& & + (-1)^{K_1+1} {K_1+K_3 \choose K_3+1} (\delta_{n0}+ \delta_{n1})
\nonumber \\
& & + (-1)^{K_1+1+n} {K_1+K_3-n-1 \choose K_3-1} 
\label{A.9}
\end{eqnarray}
\begin{eqnarray}
\Psi_{n,k} &=& (-1)^{K_1+n+1} {K_1+K_3 \choose K_3-1} \nonumber \\
& & + (-1)^{K_1+n} {K_1+K_3-n-1 \choose K_3-1}  
\label{A.10}
\end{eqnarray}
and
\begin{eqnarray}
K_1 &=& k(l_1-1) \nonumber \\
K_3 &=& k(l_3-1).
\label{A.11}
\end{eqnarray}
We have $(\alpha \in \{1,3\})$
\begin{equation}
R^{(\alpha)}_n = (-1)^n R^{(\alpha)}_2, \quad n \ge 2
\label{A.12}
\end{equation}
and
\begin{eqnarray}
(-1)^n (\Phi_{n,k} + \Phi_{n-1,k}) = (-1)^{n+1}(\Psi_{n,k} + \Psi_{n-1,k}) 
\nonumber \\
= (-1)^{K_1} {K_1+K_3-n-1 \choose K_3-2}
\label{A.13}
\end{eqnarray}
The critical coupling constants $\{g^{(2)}_k\}$ are then determined from
homogeneous equations
\begin{equation}
\sum^{l_2-1}_{k=0} (\Phi_{n,k} + \Phi_{n-1,k}) g^{(2)}_{k+1} = 0, \quad
n \ge 3
\label{A.14}
\end{equation}
with (\ref{A.13}) inserted. All $\rho^{(1)}_m, \rho^{(3)}_m$ can then be
calculated from (\ref{A.4}) except say
\[ \rho^{(3)}_{+1}, \rho^{(3)}_0, \rho^{(3)}_{-1} \]
But these follow from the first equations (\ref{A.8}) for $n \in \{0,1,2\}$
since $R^{(3)}_0, R^{(3)}_1, R^{(3)}_2$ determine these parameters uniquely. 

 Now we consider the case $(l_{1}-1)=m(l_{3}-1)$, $m\in \bbbn \backslash \{0,1\}$. Then
we have $K_{1}=mK_{3}$ and the binomial coefficients in (\ref{A.13}) have the
form
\begin{equation}
(-1)^{n}(\Phi _{n,k}+\Phi _{n-1,k})=(-1)^{mK_{3}}
{{(m+1)K_{3}-n-1} \choose {K_{3}-2}}  \label{A.15}
\end{equation}
and especially $\Phi _{2,k}$ can be expressed in terms of $%
(\Phi _{n,k}+\Phi _{n-1,k})$, $n\geq 3$:
\begin{equation}
\sum_{n=3}^{m+1}\alpha _{n} {{(m+1)K_{3}-n-1} \choose {K_{3}-2}}=
{{(m+1)K_{3}-3} \choose {K_{3}-1}}  \label{A.16}
\end{equation}
where $\alpha _{3}=\alpha _{4}=\ldots =\alpha _{m}=1$ and $\alpha _{m+1}=m+1$%
. From this we see that
\begin{equation}
R_{2}^{\left( 3\right) }=\sum_{k=0}^{l_{2}-1}g_{k+1}^{\left( 2\right) }\Phi
_{2,k}  \label{A.17}
\end{equation}
can be written as
\begin{equation}
R_{2}^{\left( 3\right) }=\sum_{i=3}^{m+1}\alpha_{i} (-1)^{i+1} \sum_{k=0}^{l_{2}-1}
g_{k+1}^{\left( 2\right) }(\Phi _{i,k} + \Phi _{i-1,k}).  
\label{A.18}
\end{equation}
Because of (\ref{A.14}) the r.h.s. of (\ref{A.18}) is zero and 
$R_{2}^{\left( 3\right) }$ is
zero, too. This means that $\rho _{-1}^{\left( 3\right) }$ is zero (see
(\ref{A.7})) which is not allowed because then the ''external'' Dyson-Schwinger
equations can not be solved.

\newpage

\setcounter{equation}{0}

\section{Maximal critical points for the (4,3,3) model}

To illustrate the fact, that for the f-matrix models with $f \geq 3$ one
can have more than one maximal critical point, we discuss the three maximal
critical points ([4,5,4], [3,3,6], [4,4,5]) of the (4,3,3) 
three-matrix model. \\[0.3cm] 

\noindent The critical point [4,5,4]: \\
The functions $r^{(i)}(z), \: i=1,2,3$ are
\begin{eqnarray}
{r_{1}}(z) &=& { \frac {1}{15}} { 
\frac {1}{z^{4}}}  - { \frac {2}{3}} 
{ \frac {1}{z^{3}}}  + { \frac {2}{z^{2
}}}  - { \frac {8}{3}} { \frac {1}{z}
}  + { \frac {5}{3}}  - { \frac {2}{5}
} z \label{B.1} \\
{r_{2}}(z) &=&  - { \frac {1}{z^{2}}}  + 
{ \frac {5}{z}}  - 10 + 10z - 5z^{2} + z^{3} \label{B.2} \\
{r_{3}}(z) &=&  - { \frac {1}{3}} { 
\frac {1}{z}}  + { \frac {7}{3}}  - { 
\frac {32}{5}} z + 9z^{2} - 7z^{3} + 3z^{4} - 
{ \frac {2}{3}} z^{5} + { \frac {1}{
15}} z^{6} \label{B.3} 
\end{eqnarray}
and we have the critical potentials
\begin{eqnarray}
{V_{1}}(x) &=& { \frac {3355}{216}} x - 
{ \frac {25}{48}} x^{2} + { \frac {
125}{8}} x^{3} - { \frac {125}{32}} x^{4} \label{B.4} \\
{V_{2}}(x) &=& { \frac {1}{2}} x^{2} + 
{ \frac {1}{45}} x^{3} \label{B.5} \\
{V_{3}}(x) &=& { \frac {72}{5}} x + { 
\frac {27}{2}} x^{2} - 3x^{3}. \label{B.6} 
\end{eqnarray}
In this case the operators $R_1$ and $-R_3$ are equal and therefore the
two commutators $[R_2, R_1]$, $[R_3, R_2]$ are obviously equal and this
critical point leads to the universality class [5,4]. \\[0,3cm]

\noindent The critical point [3,3,6]: \\
The functions $r^{(i)}(z), \: i=1,2,3$ are
\begin{eqnarray}
{r_{1}}(z) &=&  - { \frac {1}{2}} { 
\frac {\alpha }{z^{4}}}  + { \frac {13}{6}} 
{ \frac {\alpha }{z^{3}}}  - { \frac {1
}{10}} { \frac {\alpha }{z^{2}}}  - 
{ \frac {77}{10}} { \frac {\alpha }{z
}}  + { \frac {143}{15}} \alpha  - { 
\frac {17}{5}} \alpha z \label{B.7} \\
{r_{2}}(z) &=& { \frac {12}{5}} { 
\frac {\alpha }{z^{2}}}  - { \frac {26}{5}} 
{ \frac {\alpha }{z}}  + { \frac {11}{
30}} \alpha  + { \frac {61}{10}} \alpha z - 
{ \frac {9}{2}} \alpha z^{2} + { 
\frac {5}{6}} \alpha z^{3} \label{B.8} \\
{{r_{3}}(z)} &=& {{ \frac {125}{432}} 
{ \frac {\alpha }{z}}  - { \frac {18625
}{10368}} \alpha  + { \frac {8125}{1728}} \alpha
 z - { \frac {23125}{3456}} \alpha z^{2} + 
{ \frac {14375}{2592}} \alpha z^{3}} \nonumber \\ 
& & -{ {\frac {9125}{3456}} \alpha z^{4} + 
{ \frac {125}{192}} \alpha z^{5}}
 - { \frac {625}{10368}} \alpha z^{6} \label{B.9} 
\end{eqnarray}
and we have the critical potentials
\begin{eqnarray}
{V_{1}}(x) &=& { \frac {1948250}{397953}} \alpha 
x + { \frac {267493}{353736}} x^{2} + 
{ \frac {1}{3}} ({ \frac {6400}{14739
}}  + { \frac {6400}{14739}} \alpha )x^{3} \nonumber \\ 
& & + { \frac {1}{4}} ( - { \frac {1250}{
14739}} \alpha  - { \frac {625}{4913}} )x^{4} \label{B.10} \\
{V_{2}}(x) &=& { \frac {1}{2}} x^{2} + 
{ \frac {1}{3}} ( - { \frac {25}{144}
} \alpha  - { \frac {25}{144}} )x^{3} \label{B.11} \\
{V_{3}}(x) &=&  - { \frac {1393}{80}} \alpha x + 
{ \frac {26568}{625}} x^{2} + { 
\frac {1}{3}} ({ \frac {4478976}{78125}} \alpha 
 + { \frac {4478976}{78125}} )x^{3}, \label{B.12} 
\end{eqnarray}
with $\alpha$ being a solution of
\begin{equation} \label{B.13} 
{2{\alpha}^{2} + 2 {\alpha} - 1} = 0.
\end{equation}
This is one of the nonrational cases. If one proceeds to higher order
one can get higher order algebraic equations for more than one variable
in order to establish a certain critical point. This means that calculations
become very complicated or even impossible.
The operators in this case can be written as
\begin{eqnarray}
R_1 &=& R_{1,3}+a^{-\gamma}R_{1,4}+a^{-2\gamma}R_{1,5}+a^{-3\gamma}R_{1,6}
+a^{-4\gamma}R_{1,7} \label{B.14} \\
R_2 &=& R_{2,3}+a^{-\gamma}R_{2,4}+a^{-2\gamma}R_{2,5}+a^{-3\gamma}R_{2,6}
+a^{-4\gamma}R_{2,7} \label{B.15} \\
R_3 &=& R_{3,6}+a^{-\gamma}R_{3,7} \label{B.16} 
\end{eqnarray}
where the second index denotes the highest order of the differential
operators. Because of $R_{1,3}^2=R_{2,3}^2=R_{3,6}$ both commutators vanish
to leading order. But in higher orders one finds
\begin{eqnarray} \label{B.17}
\,[R_2, R_1] &=& -a^{-4\gamma} [ (\mbox{$-{\frac{25}{144}}$} \alpha-
\mbox{$\frac {169}{144}$}) \{ R_{2,3}, R_{2,4} \}+R_{1,7}-R_{2,7}, R_{2,3}] \\
\label{B.18} \,[R_3, R_2] &=& a^{-\gamma} [ \{ R_{2,3}, R_{2,4} \} + R_{3,7}, R_{2,3}].
\end{eqnarray}
During the calculation of these commutators we have used the Dyson-Schwinger 
equations which provide us with
equalities such as $R_{1,4}=R_{2,4}, R_{1,5}=R_{2,5},$ etc.. The
two commutators (\ref{B.17}, \ref{B.18}) are then commutators of an operator
of order three with one of order seven and they are equal to the operators
of the (7,3) two-matrix model, as expected. This critical point then gives
a solution that belongs to the universality class [7,3]. \\[0.3cm]

\noindent The critical point [4,4,5]: \\
The functions $r^{(i)}(z), \: i=1,2,3$ are
\begin{eqnarray}
r_{1}(z) &=& { \frac {{ 
\frac {1}{5}}  + { \frac {3}{5}} \beta  - 
{ \frac {2}{5}} \beta ^{2}}{z^{4}}}  + 
{ \frac {2\beta ^{2} - 4\beta }{z^{3}}}  + 
{ \frac { - 4\beta ^{2} - 2 + 10\beta }{z^{2}}} \nonumber \\
& & +{ \frac {4\beta ^{2} + 4 - 12\beta }{z}}  - (2
\beta ^{2} - 3 + 7\beta ) + ({ \frac {2}{5}} \beta ^{2} + 
{ \frac {4}{5}}  - { \frac {8}{5}} \beta )z \label{B.19} \\
{r_{2}}(z) &=&  - { \frac {\beta }{z^{2}}}  + 
{ \frac {4\beta  + 1}{z}}  - 6\beta  - 4 + (4
\beta  + 6)z + ( - 4 - \beta )z^{2} + z^{3} \label{B.20} \\
r_{3}(z) &=& { \frac { - 11 + 4\beta ^{2
}}{z}}  + 69 + \beta  - 26\beta ^{2} + ( - { 
\frac {926}{5}}  - { \frac {28}{5}} \beta  + 
{ \frac {362}{5}} \beta ^{2})z \nonumber \\
& & + (276 + 13\beta  - 112\beta ^{2})z^{2} + ( - 
247 - 16\beta  + 104\beta ^{2})z^{3} \nonumber \\
& & + (11\beta  - 58
\beta ^{2} + 133)z^{4} + ( - 4\beta  + 18\beta ^{2} - 40)z^{5} \nonumber \\
& & + ({ \frac {3}{5}} \beta  - { \frac {12}{5}} 
\beta ^{2} + { \frac {26}{5}} )z^{6} \label{B.21} 
\end{eqnarray}
and we have the critical potentials
\begin{eqnarray}
{V_{1}}(x) \!\!\!\!&=&\!\!\!\! { \frac {11937}{19652}} 
\beta ^{2}x \!+\! { \frac {294725}{39304}} x \!+\! 
{ \frac {263361}{39304}} \beta x  \!-\! { \frac {1}{2}} ({ \frac {8013}{
19652}} \beta ^{2} \!-\! { \frac {157853}{39304}}  \!+\! 
{ \frac {1221}{19652}} \beta )x^{2} \nonumber \\
\!\!\!\!&-&\!\!\!\! { \frac {1}{3}} \! ( { 
\frac {22959}{39304}} \beta ^{2} \!\!-\! { \frac {154635
}{19652}}  \!-\! { \frac {161649}{39304}} \beta )x^{
3} \!+\!\! { \frac {1}{4}} ({ \frac {7933}{
9826}}  \!-\! { \frac {27915}{39304}} \beta  \!-\! 
{ \frac {4254}{4913}} \beta ^{2})x^{4} \label{B.22} \\
{V_{2}}(x) \!\!\!\!&=&\!\!\!\! { \frac {1}{2}} x^{2} + 
{ \frac {1}{3}} ({ \frac {3}{5}} 
\beta  - { \frac {12}{5}} \beta ^{2} + 
{ \frac {26}{5}} )x^{3} \label{B.23} \\
{V_{3}}(x) \!\!\!\!&=&\!\!\!\! { \frac {115848}{222605}} 
\beta ^{2}x \!+\! { \frac {378696}{222605}} x \!+\! 
{ \frac {2134863}{222605}} \beta x \!+\! 
{ \frac {1}{2}} ({ \frac {21225}{
44521}}  \!+\! { \frac {219258}{44521}} \beta  \!+\! 
{ \frac {62334}{44521}} \beta ^{2})x^{2} \nonumber \\
 \!\!\!\!&+&\!\!\!\! { \frac {1}{3}} ( \!-\! { 
\frac {34521}{44521}} \beta  \!-\! { \frac {17040}{
44521}} \beta ^{2} \!-\! { \frac {4568}{44521}} )x^{3} \label{B.24} 
\end{eqnarray}
with $\beta$ being a solution of
\begin{equation} \label{B.25} 
3 \beta^3 -7 \beta -1 = 0.
\end{equation}
The operators in this case are
\begin{eqnarray}
R_1 &=& R_{1,4}+a^{-\gamma}R_{1,5} \label{B.26} \\
R_2 &=& R_{2,4}+a^{-\gamma}R_{2,5} \label{B.27} \\
R_3 &=& R_{3,5} \label{B.28} 
\end{eqnarray}
The leading order of the first commutator vanishes again ($R_{1,4}\!=\!R_{2,4}$) but the second
commmutator does not:
\begin{eqnarray} \label{B.29}
\,[R_2, R_1] &=& -a^{-\gamma} [R_{1,5}-R_{2,5}, R_{2,4}] \\
\label{B.30} \,[R_3, R_2] &=&  [ R_{3,5}, R_{2,4}].
\end{eqnarray}
Then we have $-R_{1,5}+R_{2,5}=R_{3,5}$ as operators of order five and 
find the universality class [5,4].

\newpage

\section{Examples for the three-matrix model}

In this appendix we give the potentials that belong to rational critical points
of the type that is discussed in Appendix A. This means that $\lambda_2$ assumes
its maximal possible value $(l_1+l_3-2)$. Models of the class $(5,x,3)$ were
omitted because they belong to the cases where the construction based on the
ansatz (\ref{3.5}) fails.

\setlength{\LTleft}{-18pt}
\setlength{\LTright}{\fill}

\renewcommand{\arraystretch}{1.15}

\setlongtables

\begin{longtable}{|c|c|l|} \hline

model & critical point & critical potentials \\ \hline \hline

$
3, \,2, \,3
$

&

$
[3, \,4, \,3]
$

&

$
{V_{1}} =  - 3\,x - x^{2} + { \frac {1}{3}} \,x^{3}
$ \\

& &

$
{V_{2}} = { \frac {1}{2}} \,x^{2}
$ \\

& &

$
{V_{3}} =  {V_{1}}
$ \\ \hline

$
3, \,3, \,3
$

&

$
[4, \,4, \,4]
$

&

$
{V_{1}} =  - { \frac {111}{16}} \,x + 
{ \frac {27}{8}} \,x^{2} + { \frac {3}{
4}} \,x^{3}
$ \\

& &

$
{V_{2}} = { \frac {1}{2}} \,x^{2} - { 
\frac {1}{18}} \,x^{3}
$ \\

& &

$
{V_{3}} =  {V_{1}}
$ \\ \hline

$
3, \,4, \,3
$

&

$
[5, \,4, \,5]
$

&

$
{V_{1}} =  - { \frac {1305}{112}} \,x + 
{ \frac {75}{16}} \,x^{2} + { \frac {25
}{48}} \,x^{3}
$ \\

& &

$
{V_{2}} = { \frac {1}{3}} \,x^{3} - { 
\frac {3}{140}} \,x^{4}
$ \\

& & 

$
{V_{3}} =  {V_{1}}
$\\ \hline

$
3, \,5, \,3
$

&

$
[7, \,4, \,7]
$

&

$
{V_{1}} =  - { \frac {51485}{3072}} \,x - 
{ \frac {6125}{512}} \,x^{2} + { 
\frac {1225}{768}} \,x^{3}
$ \\

& &

$
{V_{2}} = { \frac {1}{3}} \,x^{3} - { 
\frac {1}{10}} \,x^{4} + { \frac {1}{210}} \,x^{5}
$ \\

& &

$
{V_{3}} =  {V_{1}}
$ \\ \hline

$
3, \,6, \,3
$

&

$
[8, \,4, \,8]
$

&

$
{V_{1}} =  - { \frac {957285}{45056}} \,x + 
{ \frac {9261}{2048}} \,x^{2} + { 
\frac {147}{1024}} \,x^{3}
$ \\

& &

$
{V_{2}} = { \frac {1}{3}} \,x^{3} - { 
\frac {3}{4}} \,x^{4} + { \frac {9}{70}} \,x^{5} - 
{ \frac {1}{198}} \,x^{6}
$ \\

& &

$
{V_{3}} =  {V_{1}}
$ \\ \hline

$
4, \,2, \,4
$

&

$
[3, \,6, \,3]
$

&

$
{V_{1}} = { \frac {388}{27}} \,x - { 
\frac {11}{18}} \,x^{2} + { \frac {4}{27}} \,x^{3}
 + { \frac {1}{108}} \,x^{4}
$ \\

& &

$
{V_{2}} = { \frac {1}{2}} \,x^{2}
$ \\

& &

$
{V_{3}} = {V_{1}}
$ \\ \hline

$
4, \,3, \,4
$

&

$
[5, \,6, \,5]
$

&

$
{V_{1}} = { \frac {6595}{189}} \,x - 
{ \frac {125}{126}} \,x^{2} + { \frac {
250}{27}} \,x^{3} - { \frac {125}{108}} \,x^{4}
$ \\

& &

$
{V_{2}} = { \frac {1}{2}} \,x^{2} + { 
\frac {1}{105}} \,x^{3}
$ \\

& &

$
{V_{3}} = {V_{1}}
$ \\ \hline

$
4, \,4, \,4
$

&

$
[6, \,6, \,6]
$

&

$
{V_{1}} = { \frac {311127320}{5845851}} \,x + 
{ \frac {287800}{59049}} \,x^{2} + { 
\frac {160000}{59049}} \,x^{3} + { \frac {2000}{
19683}} \,x^{4}
$ \\

& &

$
{V_{2}} = { \frac {1}{2}} \,x^{2} + { 
\frac {2}{15}} \,x^{3} + { \frac {1}{528}} \,x^{4}
$ \\

& &

$
{V_{3}} = {V_{1}}
$ \\ \hline

$
4, \,5, \,4
$

&

$
[7, \,6, \,7]
$

&

$
{V_{1}} = { \frac {508518850}{6908733}} \,x - 
{ \frac {11502260}{767637}} \,x^{2} + 
{ \frac {266200}{59049}} \,x^{3} - { 
\frac {2662}{19683}} \,x^{4}
$ \\

& &

$
{V_{2}} = { \frac {1}{3}} \,x^{3} + { 
\frac {5}{264}} \,x^{4} + { \frac {1}{4862}} \,x^{5}
$ \\

& &

$
{V_{3}} = {V_{1}}
$ \\ \hline

\newpage

\hline

$
4, \,6, \,4
$

&

$
[9, \,6, \,9]
$

&

$
{V_{1}} = { \frac {300701859304}{3171108447}} \,x
 + { \frac {3004141294}{62178597}} \,x^{2} + 
{ \frac {204526784}{14348907}} \,x^{3} + 
{ \frac {7304528}{14348907}} \,x^{4}
$ \\

& &

$
{V_{2}} = { \frac {1}{3}} \,x^{3} + { 
\frac {1}{16}} \,x^{4} + { \frac {3}{1430}} \,x^{5}
 + { \frac {1}{54264}} \,x^{6}
$ \\

& &

$
{V_{3}} = {V_{1}}
$ \\ \hline

$
5, \,2, \,5
$

&

$
[3, \,8, \,3]
$

&
$
{V_{1}} =  - { \frac {24567}{400}} \,x - 
{ \frac {103}{200}} \,x^{2} - { \frac {
3}{1000}} \,x^{3} - { \frac {3}{2000}} \,x^{4} + 
{ \frac {1}{50000}} \,x^{5}
$ \\

& &

$
{V_{2}} = { \frac {1}{2}} \,x^{2}
$ \\

& &

$
{V_{3}} = {V_{1}}
$ \\ \hline

$
5, \,3, \,5
$

&

$
[5, \,8, \,5]
$

&

$
{V_{1}} =  - { \frac {109122293}{720896}} \,x - 
{ \frac {5831}{16384}} \,x^{2} + { 
\frac {232211}{24576}} \,x^{3} + { \frac {21609}{
8192}} \,x^{4} + { \frac {2401}{20480}} \,x^{5}
$ \\

& &

$
{V_{2}} = { \frac {1}{2}} \,x^{2} - { 
\frac {1}{462}} \,x^{3}
$ \\

& &

$
{V_{3}} =  {V_{1}}
$ \\ \hline

$
5, \,4, \,5
$

&

$
[7, \,8, \,7]
$

&

$
{V_{1}} = - { \frac {5926366827535}{25518145536}} \,x - 
{ \frac {500520251}{327155712}} \,x^{2} + 
{ \frac {50463788221}{981467136}} \,x^{3} $ \\
& & $ - 
{ \frac {175765205}{16777216}} \,x^{4} + 
{ \frac {35153041}{83886080}} \,x^{5}
$ \\

& &

$
{V_{2}} = { \frac {1}{2}} \,x^{2} - { 
\frac {1}{63}} \,x^{3} + { \frac {1}{19448}} \,x^{4}
$ \\

& &

$
{V_{3}} =  {V_{1}}
$ \\ \hline

$
5, \,5, \,5
$

&

$
[8, \,8, \,8]
$

&

$
{V_{1}} =  - { \frac {
318301318564336604805}{1020323700831944704}} \,x + 
{ \frac {195795357608389}{30374008717312}} \,x^{2} $ \\
& & $ + { \frac {45439661273853}{7593502179328}} \,x^{3}
+ { \frac {3691069305}{17179869184}} \,
x^{4} + { \frac {35153041}{21474836480}} \,x^{5}
$ \\

& &

$
{V_{2}} = { \frac {1}{2}} \,x^{2} - { 
\frac {3}{14}} \,x^{3} + { \frac {3}{1144}} \,x^{4}
 - { \frac {1}{148580}} \,x^{5}
$ \\

& &

$
{V_{3}} =  {V_{1}}
$ \\ \hline

$
5, \,6, \,5
$

&

$
[9, \,8, \,9]
$

&

$
{V_{1}} =  - { \frac {
489711657601739357169}{1235128690480775168}} \,x + 
{ \frac {2079610272730221}{60748017434624}} \,x^{2} $ \\
& & $ + { \frac {1697130639464183}{91122026151936}} \,x^{3} 
+ { \frac {93493458289}{137438953472}} 
\,x^{4} + { \frac {1908029761}{343597383680}} \,x^{5}
$ \\

& &

$
{V_{2}} = { \frac {1}{3}} \,x^{3} - { 
\frac {7}{572}} \,x^{4} + { \frac {7}{74290}} \,x^{5
} - { \frac {7}{35357670}} \,x^{6}
$ \\

& &

$
{V_{3}} =  {V_{1}}
$ \\ \hline

$
4, \,2, \,3
$

&

$
[3, \,5, \,3]
$

&

$
{V_{1}} = 8\,x - x^{2} + { \frac {4}{3}} \,x^{3} + 
{ \frac {1}{4}} \,x^{4}
$ \\

& &

$
{V_{2}} = { \frac {1}{2}} \,x^{2}
$ \\

& &

$
{V_{3}} = { \frac {17}{4}} \,x - { 
\frac {1}{2}} \,x^{2} - { \frac {1}{12}} \,x^{3}
$ \\ \hline

$
4, \,3, \,3
$

&

$
[4, \,5, \,4]
$

&

$
{V_{1}} = { \frac {3355}{216}} \,x - 
{ \frac {25}{48}} \,x^{2} + { \frac {
125}{8}} \,x^{3} - { \frac {125}{32}} \,x^{4}
$ \\

& &

$
{V_{2}} = { \frac {1}{2}} \,x^{2} + { 
\frac {1}{45}} \,x^{3}
$ \\

& &

$
{V_{3}} = { \frac {72}{5}} \,x + { 
\frac {27}{2}} \,x^{2} - 3\,x^{3}
$ \\ \hline

$
4, \,4, \,3
$

&

$
[5, \,5, \,5]
$

&

$
{V_{1}} = { \frac {223600}{9261}} \,x + 
{ \frac {1940}{343}} \,x^{2} + { 
\frac {8000}{1029}} \,x^{3} + { \frac {250}{343}} \,
x^{4}
$ \\

& &

$
{V_{2}} = { \frac {1}{2}} \,x^{2} + { 
\frac {4}{15}} \,x^{3} + { \frac {1}{120}} \,x^{4}
$ \\

& &

$
{V_{3}} = { \frac {29}{4}} \,x - 25\,x^{2} - 
{ \frac {25}{12}} \,x^{3}
$ \\ \hline

$
4, \,5, \,3
$

&

$
[6, \,5, \,6]
$

&

$
{V_{1}} = { \frac {4014340}{120393}} \,x - 
{ \frac {154396}{4459}} \,x^{2} + { 
\frac {53240}{1029}} \,x^{3} - { \frac {2662}{343}} 
\,x^{4}
$ \\

& &

$
{V_{2}} = { \frac {1}{3}} \,x^{3} + { 
\frac {1}{24}} \,x^{4} + { \frac {1}{1001}} \,x^{5}
$ \\

& &

$
{V_{3}} =  - 56\,x + { \frac {245}{2}} \,x^{2} - 
{ \frac {49}{3}} \,x^{3}
$ \\ \hline

\newpage

\hline

$
4, \,6, \,3
$

&

$
[7, \,5, \,7]
$

&

$
{V_{1}} = { \frac {43138744}{1008423}} \,x + 
{ \frac {1969880}{19773}} \,x^{2} + { 
\frac {2725888}{19773}} \,x^{3} + { \frac {1362944}{
59319}} \,x^{4}
$ \\

& &

$
{V_{2}} = { \frac {1}{3}} \,x^{3} + { 
\frac {9}{64}} \,x^{4} + { \frac {3}{286}} \,x^{5}
 + { \frac {1}{4896}} \,x^{6}
$ \\

& &

$
{V_{3}} =  - { \frac {927}{4}} \,x - 324\,x^{2} - 
48\,x^{3}
$ \\ \hline

$
5, \,2, \,4
$

&

$
[3, \,7, \,3]
$

&

$
{V_{1}} =  - { \frac {8331}{256}} \,x - 
{ \frac {21}{32}} \,x^{2} + { \frac {3
}{128}} \,x^{3} - { \frac {3}{128}} \,x^{4} + 
{ \frac {1}{1280}} \,x^{5}
$ \\

& &

$
{V_{2}} = { \frac {1}{2}} \,x^{2}
$ \\

& &

$
{V_{3}} =  - { \frac {626}{27}} \,x - 
{ \frac {11}{24}} \,x^{2} - { \frac {1
}{27}} \,x^{3} + { \frac {1}{864}} \,x^{4}
$ \\ \hline

$
5, \,3, \,4
$

&

$
[4, \,7, \,4]
$

&

$
{V_{1}} =  - { \frac {694729}{10000}} \,x - 
{ \frac {1029}{5000}} \,x^{2} + { 
\frac {28126}{1875}} \,x^{3} + { \frac {21609}{2500}
} \,x^{4} + { \frac {2401}{3125}} \,x^{5}
$ \\

& &

$
{V_{2}} = { \frac {1}{2}} \,x^{2} - { 
\frac {1}{210}} \,x^{3}
$ \\

& &

$
{V_{3}} =  - { \frac {26045}{378}} \,x - 
{ \frac {1375}{126}} \,x^{2} - { 
\frac {125}{9}} \,x^{3} - { \frac {125}{108}} \,x^{4
} 
$ \\ \hline

$
5, \,4, \,4
$

&

$
[5, \,7, \,5]
$

&

$
{V_{1}} =  - { \frac {239105243531}{2190240000}} \,
x - { \frac {4369673}{780000}} \,x^{2} + 
{ \frac {10962726929}{63180000}} \,x^{3} $ \\
& & $ - { \frac {35153041}{405000}} \,x^{4} + 
{ \frac {35153041}{4050000}} \,x^{5}
$ \\

& &

$
{V_{2}} = { \frac {1}{2}} \,x^{2} - { 
\frac {2}{63}} \,x^{3} + { \frac {1}{4576}} \,x^{4}
$ \\

& &

$
{V_{3}} =  - { \frac {12680}{77}} \,x + 
{ \frac {2144}{7}} \,x^{2} - { \frac {
512}{3}} \,x^{3} + 16\,x^{4}
$ \\ \hline

$
5, \,5, \,4
$

&

$
[6, \,7, \,6]
$

&

$
{V_{1}} =  - { \frac {140287221613}{940970706}} \,x
 + { \frac {665951209}{8739666}} \,x^{2} + 
{ \frac {80349808}{336141}} \,x^{3} $ \\
& & $+ { \frac {175765205}{3084588}} \,x^{4} + 
{ \frac {35153041}{11567205}} \,x^{5}
$ \\

& &

$
{V_{2}} = { \frac {1}{2}} \,x^{2} - { 
\frac {3}{14}} \,x^{3} + { \frac {3}{572}} \,x^{4}
 - { \frac {1}{35530}} \,x^{5}
$ \\

& &

$
{V_{3}} =  - { \frac {326419}{378}} \,x - 
{ \frac {14641}{14}} \,x^{2} - { 
\frac {6655}{27}} \,x^{3} - { \frac {1331}{108}} \,x
^{4}
$ \\ \hline

$
5, \,6, \,4
$

&

$
[7, \,7, \,7]
$

&

$
{{V_{1}} =  - { \frac {2615115300936941}{
14045490038528}} \,x - { \frac {7713352643381}{
85875958116}} \,x^{2} + { \frac {98261624661739}{
2748030659712}} \,x^{3}} $ \\ 
& & $- { \frac {4581179456161}{2748030659712
}} \,x^{4} + { \frac {4581179456161}{247322759374080
}} \,x^{5}\mbox{\hspace{132pt}}
$ \\

& &

$
{V_{2}} = { \frac {1}{2}} \,x^{2} - { 
\frac {16}{7}} \,x^{3} + { \frac {45}{286}} \,x^{4}
 - { \frac {4}{1615}} \,x^{5} + { 
\frac {1}{91770}} \,x^{6}
$ \\

& &

$
{V_{3}} =  - { \frac {1220198}{297}} \,x + 
{ \frac {62209}{72}} \,x^{2} - { 
\frac {343}{9}} \,x^{3} + { \frac {343}{864}} \,x^{4
}
$ \\ \hline

$
6, \,2, \,3
$

&

$
[3, \,7, \,3]
$

&

$
{V_{1}} = 24\,x - x^{2} + 3\,x^{4} + { \frac {8}{5}
} \,x^{5} + { \frac {1}{6}} \,x^{6}
$ \\

& &

$
{V_{2}} = { \frac {1}{2}} \,x^{2}
$ \\

& &

$
{V_{3}} = { \frac {29}{4}} \,x - { 
\frac {1}{4}} \,x^{2} - { \frac {1}{48}} \,x^{3}
$ \\ \hline

$
6, \,3, \,3
$

&

$
[4, \,7, \,4]
$

&

$
{V_{1}} = { \frac {4217031}{100000}} \,x - 
{ \frac {729}{1600}} \,x^{2} + { 
\frac {40743}{400}} \,x^{3} - { \frac {308367}{320}
} \,x^{4} $ \\
& & $ + { \frac {177147}{160}} \,x^{5} - 
{ \frac {19683}{64}} \,x^{6}
$ \\

& &

$
{V_{2}} = { \frac {1}{2}} \,x^{2} + { 
\frac {1}{135}} \,x^{3}
$ \\

& &

$
{V_{3}} =  - { \frac {423}{5}} \,x + 
{ \frac {27}{2}} \,x^{2} - 3\,x^{3}
$ \\ \hline

$
6, \,4, \,3
$

&

$
[5, \,7, \,5]
$

&

$
{V_{1}} = { \frac {1157614349283}{
18564650000}} \,x - { \frac {6163075107}{11881376000
}} \,x^{2} + { \frac {20496602637}{14851720}} \,x^{3
} $ \\ 
& & $ + { \frac {4321777591467}{475255040}} \,x^{4} + { \frac {80387359983}{7425860}} \,x^{
5} + { \frac {80387359983}{23762752}} \,x^{6}
$ \\

& &

$
{V_{2}} = { \frac {1}{2}} \,x^{2} + { 
\frac {4}{81}} \,x^{3} + { \frac {1}{1680}} \,x^{4}
$ \\

& &

$
{V_{3}} =  - { \frac {44505}{28}} \,x - 675\,x^{2}
 - { \frac {675}{4}} \,x^{3}
$ \\ \hline

\newpage

\hline

$
6, \,5, \,3
$

&

$
[6, \,7, \,6]
$

&

$
{V_{1}} = { \frac {34173103071667}{
408422300000}} \,x - { \frac {18059440511}{
32673784000}} \,x^{2} + { \frac {3311338930137}{
1633689200}} \,x^{3} $ \\
& & $ - { \frac {684760877857}{
118813760}} \,x^{4} + { \frac {41615795893}{11881376}} \,x
^{5} - { \frac {41615795893}{71288256}} \,x^{6}
$ \\

& &

$
{V_{2}} = { \frac {1}{2}} \,x^{2} + { 
\frac {1}{3}} \,x^{3} + { \frac {1}{70}} \,x^{4} + 
{ \frac {1}{7315}} \,x^{5}
$ \\

& &

$
{V_{3}} =  - 10695\,x + { \frac {6125}{2}} \,x^{2}
 - { \frac {1225}{3}} \,x^{3}
$ \\ \hline

$
6, \,6, \,3
$

&

$
[7, \,7, \,7]
$

&

$
{{V_{1}} = { \frac {788322756540293184}{
7468028442184375}} \,x \! + \! { \frac {3521389972070481}{
298721137687375}} \,x^{2} \! + \! { \frac {865581376911776
}{2757425886345}} \,x^{3}} $ \\ 
& & $ + { \frac {1969585683008454}{
11948845507495}} \,x^{4} + { \frac {242703321647976
}{11948845507495}} \,x^{5} + { \frac {3370879467333
}{4779538202998}} \,x^{6}
$ \\

& &

$
{V_{2}} = { \frac {1}{2}} \,x^{2} + { 
\frac {32}{9}} \,x^{3} + { \frac {3}{7}} \,x^{4} + 
{ \frac {8}{665}} \,x^{5} + { \frac {1
}{10530}} \,x^{6}
$ \\

& &

$
{V_{3}} =  - { \frac {180675}{4}} \,x - 
{ \frac {11907}{4}} \,x^{2} - { \frac {
1323}{16}} \,x^{3}
$ \\ \hline

$
6, \,2, \,4
$

&

$
[3, \,8, \,3]
$

&

$
{V_{1}} = { \frac {190536}{3125}} \,x - 
{ \frac {851}{1250}} \,x^{2} - { 
\frac {16}{1875}} \,x^{3} + { \frac {1}{50}} \,x^{4} $ \\
& & $ + { \frac {8}{3125}} \,x^{5} + { 
\frac {1}{18750}} \,x^{6}
$ \\

& &

$
{V_{2}} = { \frac {1}{2}} \,x^{2}
$ \\

& &

$
{V_{3}} = { \frac {174}{5}} \,x - { 
\frac {3}{8}} \,x^{2} + { \frac {1}{75}} \,x^{3} + 
{ \frac {1}{4000}} \,x^{4}
$ \\ \hline

$
6, \,3, \,4
$

&

$
[4, \,8, \,4]
$

&

$
{{V_{1}} = }  { \frac {775352871324}{6355146875}} \,x - 
{ \frac {3256929}{20336470}} \,x^{2} + 
{ \frac {41065704}{4621925}} \,x^{3} - 
{ \frac {52337097}{1848770}} \,x^{4} $ \\
& & $ + { \frac {708588}{84035}} \,x^{5} - { 
\frac {19683}{33614}} \,x^{6}
$ \\

& &

$
{V_{2}} = { \frac {1}{2}} \,x^{2} + { 
\frac {4}{1485}} \,x^{3}
$ \\

& &

$
{V_{3}} = { \frac {5825}{27}} \,x - { 
\frac {4325}{24}} \,x^{2} + { \frac {125}{2}} \,x^{3
} - { \frac {125}{32}} \,x^{4}
$ \\ \hline

$
6, \,4, \,4
$

&

$
[5, \,8, \,5]
$

&

$
{V_{1}} = { \frac {45952369797}{244268750}} \,x - 
{ \frac {986577}{1645600}} \,x^{2} + { 
\frac {2048274}{4675}} \,x^{3} + { \frac {859977}{
880}} \,x^{4} $ \\ 
& & $ + { \frac {1944}{5}} \,x^{5} + { \frac {81}{2}} \,x^{6}
$ \\

& &

$
{V_{2}} = { \frac {1}{2}} \,x^{2} + { 
\frac {8}{495}} \,x^{3} + { \frac {1}{15504}} \,x^{4
}
$ \\

& &

$
{V_{3}} = { \frac {169600}{27}} \,x + 7550\,x^{2}
 + { \frac {8000}{3}} \,x^{3} + 250\,x^{4}
$ \\ \hline

$
6, \,5, \,4
$

&

$
[6, \,8, \,6]
$

&

$
{{V_{1}} = { \frac {2294017899932829}{
8980595525000}} \,x - { \frac {30776961716517}{
1436895284000}} \,x^{2} + { \frac {1636161080121}{
780921350}} \,x^{3}} $ \\
& & $ - { \frac {320385186351}{113588560}} \,
x^{4} + { \frac {1203384114}{1419857}} \,x^{5} - 
{ \frac {200564019}{2839714}} \,x^{6}
\mbox{\hspace{82pt}}
$ \\

& &

$
{V_{2}} = { \frac {1}{2}} \,x^{2} + { 
\frac {4}{45}} \,x^{3} + { \frac {1}{816}} \,x^{4}
 + { \frac {6}{1562275}} \,x^{5}
$ \\

& &

$
{V_{3}} = { \frac {8025008200}{120393}} \,x - 
{ \frac {227667550}{4459}} \,x^{2} + { 
\frac {13310000}{1029}} \,x^{3} - { \frac {332750}{
343}} \,x^{4}
$ \\ \hline

$
6, \,6, \,4
$

&

$
[7, \,8, \,7]
$

&

$
{V_{1}} \! = \! { \frac {154075511895816746448}{
476632759170528125}} \,x \! + \! { \frac {
23090824363335471}{131484899081525}} \,x^{2} \! + \! { 
\frac {9611096332631616}{5716734742675}} \,x^{3} $ \\
& & $ + { \frac {928245980327382}{
1143346948535}} \,x^{4} + { \frac {112290178445568}{
1143346948535}} \,x^{5} + { \frac {779792905872}{
228669389707}} \,x^{6}
$ \\

& &

$
{V_{2}} = { \frac {1}{2}} \,x^{2} + { 
\frac {16}{27}} \,x^{3} + { \frac {3}{136}} \,x^{4}
 + { \frac {24}{120175}} \,x^{5} + { 
\frac {1}{1941840}} \,x^{6}
$ \\

& &

$
{V_{3}} = { \frac {10068806}{27}} \,x + 
{ \frac {2546203}{24}} \,x^{2} + 10648\,x^{3} + 
{ \frac {1331}{4}} \,x^{4}
$ \\ \hline

\end{longtable}

\end{appendix}

\end{document}